# Highly efficient polarization-entangled photon-pair generation in lithium niobate waveguides based on bound states in continuum

**G.Y. CHEN, Z.X. LI, Y.H CHEN, AND X.D. ZHANG***

*Key Laboratory of advanced optoelectronic quantum architecture and measurements of Ministry of Education, Beijing Key Laboratory of Nanophotonics & Ultrafine Optoelectronic Systems, School of Physics, Beijing Institute of Technology, 100081, Beijing, China*

** zhangxd@bit.edu.cn*

**Abstract :** Integrated optics provides a platform for the experimental implementation of highly complex and compact circuits for practical applications as well as for advances in the fundamental science of quantum optics. The lithium niobate (LN) waveguide is an important candidate for the construction of integrated optical circuits. Based on the bound state in the continuum (BIC) in a LN waveguide, we propose an efficient way to produce polarization-entangled photon pairs. The implementation of this method is simple and does not require the polarization process needed for periodically poled LN. The generation rate of the entangled photon pairs increases linearly with the length of the waveguide. For visible light, the generation efficiency can be improved by more than five orders of magnitude with waveguides having the length of only a few millimeters, compared with the corresponding case without BICs. The phenomena can appear in a very wide spectrum range from the visible to THz regions. This study is of great significance for the development of active integrated quantum chips in various wavelength ranges.

**Keywords**: entangled photon pair, lithium niobate waveguide, bound state in continuum

## 1. Introduction

In recent decades, there has been a great interest in the production of entangled photon pairs because they are an essential resource for many novel functionalities in quantum information processing [1–4]. Various approaches have been used to produce such entangled pairs [5–20]. A well-known approach is based on using nonlinear crystals such as β-bariumborate (BBO) for nonlinear parametric down-conversion [5]. Other methods, such as the use of quantum dots, quasi-phase matching in photonic crystals, periodically poled materials, and two-dimensional materials have also been proposed theoretically [6-15] and demonstrated experimentally [16-20]. However, these methods mainly focus on the production of entangled photons in free space.

Recently, the approach of integrated optics is regarded to be essential for practical applications as well as for advances in the fundamental science of quantum optics [21–23]. The integration of various optical components on the same chip allows the realization of on-chip optical gate operations [24], multi-photon quantum interference [25], quantum Boson sampling [26], and the simulation of quantum walks [27]. An essential aspect for implementing a fully integrated platform for quantum information processing requires the realization of quantum light sources, in particular, entangled photon sources, on a chip. Although some schemes have been proposed [24], there is still a lack of optimal designs for on-chip generation of entangled photons.

Among various integrated platforms, lithium niobate (LN) is an attractive candidate for the development of advanced integrated quantum circuits. Not only does its possess outstanding electro-optic, acousto-optic, and nonlinear optical properties [28–30], but it is also compatible with modern fabrication techniques. For example, photon-pair generation based on quasi-phase matching in periodically poled LN waveguides has undergone rapid development and shown promising conversion efficiencies [31–46].

In this work, we propose a new way to efficiently produce polarization-entangled photon pairs by designing bound states in continuum (BICs). Analogous to localized electrons with energies larger than their potential barriers, light BICs, which are also known as embedded trapped modes and correspond to discrete eigenvalues which coexist with the extended modes of a continuous spectrum, have been realized in recent years [47–58]. They have been shown to exist in dielectric gratings, waveguide structures, object surfaces, photonic crystal slabs, and some open subwavelength nanostructures [59–68]. Recently, photonic integrated circuits with BICs have been designed using the LN platform [69, 70]. Here, we show theoretically the improvement of generation rate for the polarization-entangled photon pair by several orders of magnitude in LN waveguides with BICs, compared with the corresponding case without BICs. This design does not require the dedicate polarization process necessary for periodically poled LN waveguides. The phenomenon can appear in a wide spectral range from the visible to the THz regions.

**2. System and theory**

We consider a waveguide structure composed of a z-cut LN layer, which optical axis along the z axis [69], sandwiched between a low-refractive index organic polymer layer and a silica ($SiO_2$) substrate as shown in Fig. 1(a) and Fig. 1(b). In Refs. [69, 70], it was proposed that such a structure supports a transverse-magnetic (TM) bound mode which lies within the continuous spectrum of the transverse-electric (TE) modes when the parameters are properly selected. The effective refractive indices of different waveguide modes in the structure can be calculated by the planar waveguide analysis method. The propagation of electromagnetic field in the waveguide structure satisfies the wave equation. Using the boundary condition of field continuity in the structure, we can calculate wave propagation constants. Then, the effective refractive indices of different waveguide modes can be obtained. Because the LN is an anisotropic material, we usually need to calculate the effective indices along different directions simultaneously. However, for a z-cut LN layer, the optical axis is along the z axis. It can be demonstrated that non-zero terms are only diagonal elements [77]. In such a case, the effective indices along different directions can be calculated separately. The details of the calculation method are given in Refs.[69, 70].

In Fig. 1(c), we present the distributions of the effective refractive indices as functions of the spatial position y. Here, the following parameters are taken: the pump wavelength $\lambda_p = 633nm,$ the thickness of the silica slab $d = 2.1\mu m,$ the thickness of the organic waveguide $h_{wg} = 490nm,$ the thickness of the LN waveguide $d_{LN} = 380nm,$ and the width of the waveguide $w = 1.6\mu m.$ The refractive indices of silica, the polymer, and LN are $n_{SiO_2} = 1.45, n_{wg} = 1.54287$ and $n_{LN-p} = 2.2864,$ respectively [69]. As shown in Fig.1(c), the effective refractive indices of TE and TM modes inside the waveguide are 2.182 and 2.145, respectively. They become 2.179 and 2.135 outside the waveguide.

It is easy to find that the potential distribution for photons is inversely related to the refractive index distribution by comparing the Schrödinger equation with the wave equation [69,70]. Thus, we can calculate the photonic potential distributions for the above waveguide structure after obtaining the effective refractive indices. Fig. 1(d) (only a schematic) shows the distributions of the structure potential well for TM and TE modes as functions of the spatial position y. It is clear that the TM potential well can support a TM bound mode, which lies in the spectrum of the TE continuous modes (only two discrete orders of continuous TE modes are shown in Fig. 1(d) for illustration).

In fact, the low-refractive index material is patterned in the lateral direction (y direction) and forms a high-effective index channel for the transverse confinement of the TE and TM modes. This makes the TM mode bound within the LN waveguide and the TE mode extended. The spatial confinement of the TE and TM modes can also be analyzed by the transmission loss spectrum and the electric field distribution in the structure. According to the analysis in Ref. [69], the transmission loss can be described by $I(x)/I(0) = \exp(-x/L_d),$ where $I(0)$ is the initial intensity, and $I(x)$ is the intensity at the position $x$. Here $L_d = \frac{w^2}{\sin^2\left(k_y w/2\right)} \frac{\sqrt{n_{TE,2}^2 - n_{eff-p}^2}}{n_{eff-p}^2}$ represents the decay length, $k_y = \sqrt{n_{TE,1} - n_{eff-p}^2} k_0$, $k_0$ is the wave vector in vacuum, $n_{eff-p}$ is the effective refractive index for the pump field, $n_{TE,1}$ and $n_{TE,2}$ represent effective indices with and without polymer waveguide, respectively. We can design structural parameters so that $\sin^2\left(k_y w/2\right) = 0$. In such a case, $L_d$ becomes infinite and $I(x) = I(0)$, the dissipation can be avoided. Fig. 1(e) shows the transmission loss as a function of the waveguide width, and the inset displays the result of local magnification from $w = 1.5\ \mu m$ to $w = 2\ \mu m$ with logarithm axis. The zero loss appears at $w = 1.6\ \mu m$. In Fig. 1(f), we plot the electric field distribution for the TM mode. The corresponding y and z components are given in Figs. 1(g) and (h), respectively. These results are obtained by the finite element method (COMSOL Mutiphysics). It is seen that almost all the fields are concentrated in the LN waveguide, at the same time, there is no propagation loss. In such a case, the TE mode is in the region of the continuous spectrum, and the system shows good characteristics of BICs.

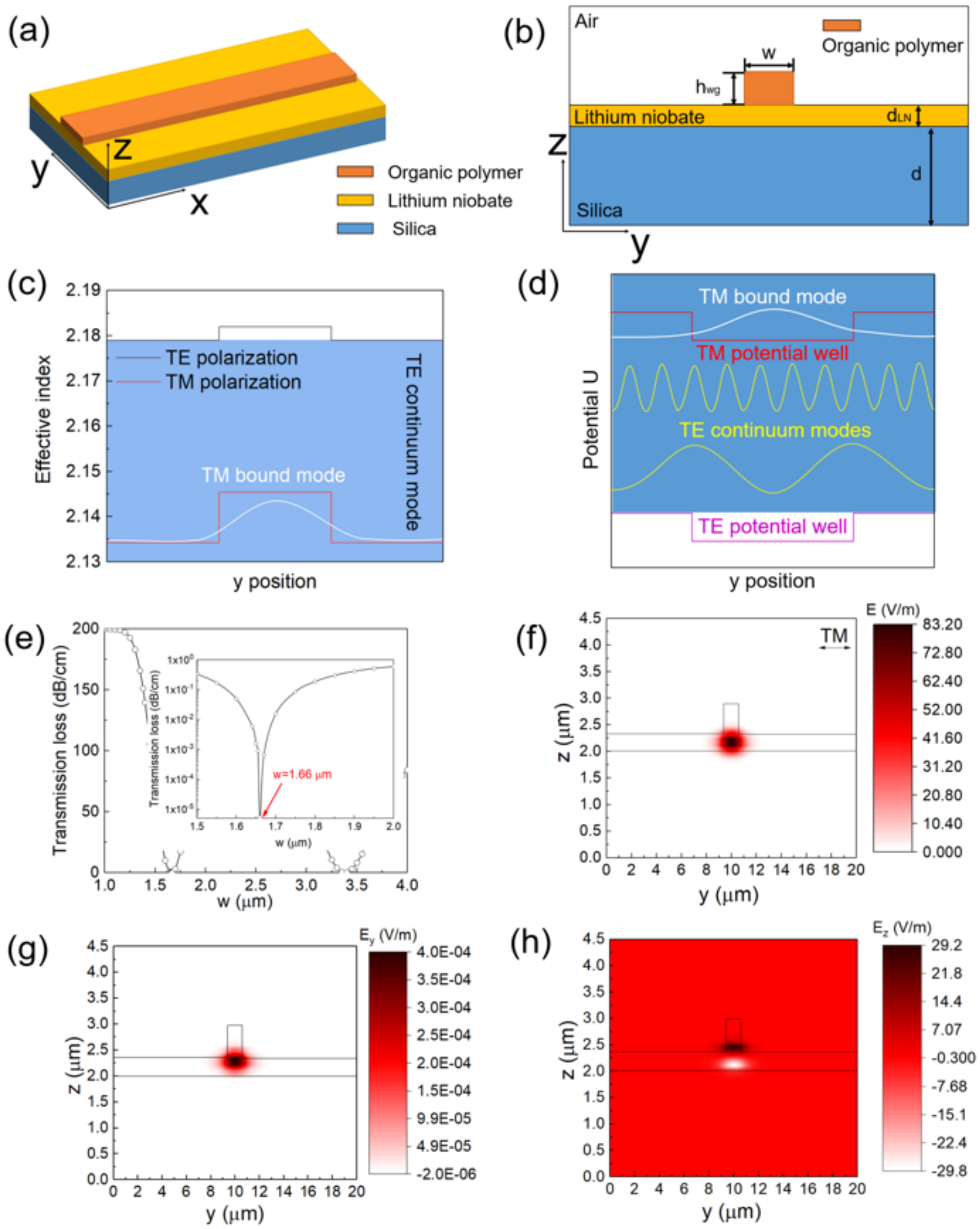

Fig. 1. (a) The BIC structure consists of a polymer layer, a z-cut LN layer and a silica substrate. (b) Cross section of the waveguide. (c) The effective refractive index distribution along the y axis. The black line represents the distribution of TE modes in and outside the waveguide, the red line represents the distribution of TM modes in and outside the waveguide, the blue part represents the continuum TE mode, and the white line represents the fitted BIC mode distribution. (d) Potential well distributions of the TM and TE modes. The potential well for TM modes is higher than that of TE modes and lies within the continuum TE modes. (e) The relation between the transmission loss as a function of the width of the waveguide. The inset shows the transmission loss as a function of width of the waveguide from $w = 1.5\ \mu m$ to $w = 2\ \mu m$. The red arrow indicates that the transmission loss is 0 at $w = 1.6\ \mu m$. (f) The corresponding electric field distribution for the TM mode, (g) and (h) for y and z components, respectively.

We now discuss the generation of coherent photon pairs in the above structure. The system is divided into the three layers of air-LN-air along the direction of propagation. The nonlinear interaction in the structure is described by the Hamiltonian $H(t)$ [71]:

$$H(t) = \varepsilon_0 \int_V \chi(\vec{\boldsymbol{r}}) * \left[ E_p(r,t) * E_s^{\dagger}(r,t) * E_i^{\dagger}(r,t) + h.c. \right], \tag{1}$$

where $\varepsilon_0$ is the vacuum dielectric constant, $V$ is the volume of the LN waveguide, and $\chi(\vec{\boldsymbol{r}})$ is the LN second-order nonlinear coefficient. $E_p(r,t)$, $E_s^{\dagger}(r,t)$ and $E_i^{\dagger}(r,t)$ represent the pump light field, signal light field, and the idler light field, respectively. The h.c. stands for the Hermitian conjugated term. In Ref. [71], the quantum theory for the generation of correlated photon pairs from the heterostructure has been presented for the case of plane wave incidence. Now, we extend the original theory to the waveguide structures with finite size. Considering the monochromatic pump field incident on the waveguide structures, $E_s^{\dagger}(r,t)$ and $E_i^{\dagger}(r,t)$ can be written as:

$$E_m(r,t)=\frac{1}{2}\left(\mathrm{E}_m\left(\vec{r},t\right)+\mathrm{E}_m^{\dagger}\left(\vec{r},t\right)\right)=\frac{i}{2}\left(\sqrt{\frac{2\hbar\omega_m}{n_m^2\varepsilon_0}}*U_m\left(\vec{r}\right)*e^{-i\omega_m t}*a_m+h.c.\right)(m=s,i), \quad (2)$$

where the $\omega_m\ (m=s,i)$ represent the angular frequencies of the signal light (idler light). The $n_m\ (m=s,i)$ are the refractive indices of the material for the signal light and idler light. The $\alpha_m\ (m=s,i)$ represent annihilation operators. In Eq.(2), the location-dependent functions $U_m\left(\vec{r}\right)\ (m=s,i)$ are introduced to describe the transmission behavior in the waveguide system, which meet the following conditions:

$$\begin{cases}\left(\nabla^2+\dfrac{\omega_m^2}{n^2c^2}\right)U_m=0\\ \nabla\bullet U_m=0\\ \int_V U_m^*U_m=1\end{cases}, \quad (3)$$

The last integral depends on the waveguide volume $V$. We assume that in the waveguide the solution can be written as:

$$U_m\left(\vec{r}\right)=\frac{1}{\sqrt{L}}U_m(x,y)e^{i\beta_m z}, \quad (4)$$

where $L$ presents the length of the LN waveguide, $\beta_m$ is the propagation constant of signal (idler) light, $U_m(x,y)$ satisfies $\int_A dxdy\left|U_m(x,y)\right|^2=1$, and $A$ is the effective cross-sectional area of the waveguide. Similarly, the location-dependent function $U_p(x,y)$ for the pump field can also be introduced to describe the transmission behavior in the waveguide systems, which also satisfies $\int dxdy\left|U_p(x,y)\right|^2=1$ and $P_p$ is the power of the pump field.

In such a case, $E_p(r,t)$ can be expressed as

$$E_p(\vec{r},t)=\frac{1}{2}\left(\mathrm{E}_p(\vec{r},t)+\mathrm{E}_p^{\dagger}(\vec{r},t)\right)=\frac{1}{2}\left(\sqrt{\frac{2P_p}{cn_p\varepsilon_0}}*U_p(x,y)*e^{i(\beta_p z-\omega_p t)}+h.c.\right), \quad (5)$$

where $c$ is the speed of light, $\beta_p$ is the propagation constant under the pump light, $\omega_p$ is the angular frequency of the pump light, $n_p$ is the refractive index of the material, $\hbar$ is the reduced Planck constant. After substituting Eqs. (2), (3), (4) and (5) into Eq. (1), the Hamiltonian can be expressed as:

$$\begin{aligned}H(t)=&-\frac{d\hbar\sqrt{2}}{\sqrt{\varepsilon_0 c n_s^2 n_i^2 n_p}}*\frac{\sqrt{P_p}}{L}\\&*\sqrt{\omega_s\omega_i}\int_V dxdydz\left[U_pU_s^*U_i^* * e^{i(\beta_p-\beta_s-\beta_i)z}*e^{i(\omega_p-\omega_s-\omega_i)t}*a_s^{\dagger}a_i^{\dagger}+h.c.\right],\end{aligned} \quad (6)$$

Here we assume that only the intense pump wave is fed into the sample and the initial state is the vacuum state $|vac\rangle$. Solving the Schrodinger equation to first order in the nonlinear perturbation, the output states $|\Psi\rangle_{s,i}^{out}$ of the signal and idler fields can be expressed as [71]:

$$\begin{aligned}|\Psi\rangle_{s,i}^{out}&=\exp\left[-\frac{i}{\hbar}\int_{-\infty}^{\infty}dtH(t)\right]|vac\rangle\\&=|vac\rangle-\frac{i}{\hbar}\lim_{T\to\infty}\int_{-T}^{T}dtH(t)|vac\rangle,\end{aligned} \quad (7)$$

In the above equation, only LN has a nonzero second-order nonlinear coefficient and a negligible propagation loss. Then, the output state can be written as:

$$|\Psi\rangle_{s,i}^{out}=|vac\rangle-\frac{id\sqrt{2}}{\sqrt{\varepsilon_0 c n_s^2 n_i^2 n_p}}*\frac{\sqrt{P_p}}{L}*\sqrt{\omega_s\omega_i}\int_V dxdydz\left[U_pU_s^*U_i^* * e^{i\Delta\beta z}*a_s^{\dagger}a_i^{\dagger}+h.c.\right]|vac\rangle, \quad (8)$$

We are interested only in the second term on the right side of Eq. (8). The vacuum state $|vac\rangle$ is thus discarded. We consider the part where both the photons are emitted in the forward direction, that is, the signal light and the idler light are emitted along the same direction as the incident pump light (Front-Front, denoted as “FF”):

$$|\Psi\rangle_{s,i}^{out}=|\Psi\rangle_{s,i}^{FF}=\int_0^{\infty}d\omega_s\int_0^{\infty}d\omega_i\left[\phi^{FF}(\omega_s,\omega_i)*\alpha_s^{\dagger}\alpha_i^{\dagger}(\omega_i)\right]|vac\rangle, \quad (9)$$

where $\phi^{FF}(\omega_s,\omega_i)$ represents the probability amplitude that a photon pair occurs in modes signal-forward and idler-forward, and can be written as:

$$\phi^{FF}(\omega_s,\omega_i)=\sum_{\substack{c=11\\c=21}}\sum_{\substack{d=11\\d=21}}\Gamma*\left(F_{sF}^{(2)}\right)_c(\omega_s)*\left(F_{iF}^{(2)}\right)_d(\omega_i), \tag{10}$$

with

$$\Gamma=\sum_{\substack{b=11\\b=21}}-\frac{id\sqrt{2}}{\sqrt{\varepsilon_0 c n_s^2 n_i^2 n_p}}*\frac{\sqrt{P_p}}{L}*\sqrt{\omega_s\omega_i}*\left(F_{pF}^{(2)}\right)_b(\omega_p)*\int_V dxdydzU_pU_s^{\dagger}U_i^{\dagger}e^{i\Delta kz}, \tag{11}$$

The explicit expressions of $\left(F_{sF}^{(2)}\right)_c$, $\left(F_{iF}^{(2)}\right)_d$, and $\left(F_{pF}^{(2)}\right)_b$ are given in the Appendix, where the subscripts b, c and d are the indices of the matrix elements. Based on the relation $\int_A dxdy\left|U_m(x,y)\right|^2=1$, the integral in Eq.(11) can be expressed as:

$$\int_V dxdydzU_pU_s^*U_i^*e^{i\Delta kz}=L*A*\sin c\left(\frac{\Delta k*L}{2}\right), \tag{12}$$

The square of the probability amplitude $\phi^{FF}(\omega_s,\omega_i)$ (probability density) can be written as:

$$\left|\phi^{FF}(\omega_s,\omega_i)\right|^2=\lim_{T\to\infty}2T*f(\omega_s,\omega_i)\delta(\omega_p-\omega_s-\omega_i), \tag{13}$$

with

$$f(\omega_s,\omega_i)=|\varphi|^2, \tag{14}$$

and

$$\varphi=\sum_{\substack{c=11\\c=21}}\sum_{\substack{d=11\\d=21}}\Gamma*\left(F_{sF}^{(2)}\right)_c(\omega_s)*\left(F_{iF}^{(2)}\right)_d(\omega_i). \tag{15}$$

After the output states are obtained, the generation rate of correlated photon pairs can be calculated. We define a quantity $N_{s,i}^{FF}(\omega_s,\omega_i)$ which describes the counts of photon pairs that have a signal photon of frequency $\omega_s$ and an idler photon of frequency $\omega_i$. By using Eq. (7) and Eq. (13), the expression $N_{s,i}^{FF}(\omega_s,\omega_i)$ can be written as [71]:

$$\begin{aligned}N_{s,i}^{FF}(\omega_s)&=\langle\Psi|_s^{out}a_s^{\dagger}(\omega_s)a_s(\omega_s)a_i^{\dagger}(\omega_i)a_i(\omega_i)|\Psi\rangle_i^{out}=\int_0^\infty d\omega_i\left|\phi^{FF}(\omega_s,\omega_i)\right|^2\\&=\left(\sum_{\substack{c=11\\c=21}}\sum_{\substack{d=11\\d=21}}\left(F_{sF}^{(2)}\right)_c(\omega_s)*\left(F_{iF}^{(2)}\right)_d(\omega_i)\right)^2\\&*\left(\sum_{\substack{b=11\\b=21}}-\frac{id\sqrt{2}}{\sqrt{\varepsilon_0 c n_s^2 n_i^2 n_p}}*\frac{\sqrt{P_p}}{L}*\sqrt{\omega_s\omega_i}*\left(F_{pF}^{(2)}\right)_b(\omega_p)*L*A*\sin c\left(\frac{\Delta k*L}{2}\right)\right)^2,\end{aligned} \tag{16}$$

$$N_s^{FF}=N_i^{FF}=f(\omega_s,\omega_p-\omega_s)=f(\omega_i,\omega_p-\omega_i). \tag{17}$$

We are now ready to use the power of the signal field $P_s$ to measure the conversion efficiency of correlated photon pairs. In the one-dimensional state space, the relationship between the change of signal photon number and the change of energy is [72]:

$$dN = \frac{L}{2\pi} * \frac{n_s}{\hbar c} dE, \tag{18}$$

The same holds for the idler light. Thus, the density of state (DOS) is given by [72]:

$$\rho = \frac{L^2}{(2\pi)^2} * \frac{n_s n_i}{\hbar c^2} d\omega_s, \tag{19}$$

The above formula can be extended to three-dimensional space by using Fermi Golden Rule [72]:

$$W = \frac{2\pi}{\hbar} * \left|\phi^{FF}(\omega_s, \omega_i)\right|^2 * \rho, \tag{20}$$

The differential of the down-converted signal optical power can be expressed as:

$$dP^{(W)}{}_s = \hbar\omega_s * W, \tag{21}$$

Then, the power at the different signal wavelength is obtained:

$$dP_s = \hbar * \frac{2\pi}{\hbar} * \left|\phi^{FF}(\omega_s, \omega_i)\right|^2 * \rho * d\omega_s, \tag{22}$$

Using Eqs. (17) and (22), the mean number of photon pairs and the conversion efficiency can be obtained easily by numerical calculations.

## 3. Numerical results and discussions

In the following, we present numerical results for the generation of polarization-entangled photon pairs based on the LN waveguide. The technique for numerical calculations is not difficult, which can be realized by using MATLAB or any computation language such as Fortran and so on. We first consider the structure parameters as shown in Fig. 1, where the wavelength of the pump is $\lambda$=633$nm$, and the nonlinear coefficient of LN is $d_{33} = 27 pm/V$ [75]. Fig. 2 shows the calculated number of coherent photon pairs as a function of the waveguide length. As the length increases, the number of coherent photon pairs produced increases in an oscillating manner. The oscillation period is determined by $\Delta k * L$, as shown in Eq. (16). Here $\Delta k$ depends on the quasi-phase matching conditions, that is, for entangled photon pairs, the conservation of energy and momentum needs to be satisfied: $\omega_p = \omega_s + \omega_i$ and $k_p = k_s + k_i$. If $\Delta k = k_p - k_s - k_i$ is zero, then there is no oscillation. In fact, the perfect phase matching condition is difficult to realize in real systems. For example, in the present waveguide system, $\Delta k$ is expressed as [72]:

$$\Delta k = 2\pi * \left( \frac{n_{eff-p}}{\lambda_p} - \frac{n_{eff-s}}{\lambda_s} - \frac{n_{eff-i}}{\lambda_i} \right), \tag{23}$$

where $n_{eff-s}$ and $n_{eff-i}$ represent the effective refractive indices at the wavelengths of the signal and idler lights, respectively. These effective refractive indices can be obtained by calculating the propagation constants of the corresponding waves in the structure. The results shown in Fig.2 are the self consistent solution of Eq. (23) that makes $\Delta k$ minimum. In such a case, $\Delta k = 2.1*10^{-4}\ nm^{-1}$, $n_{eff-p} = 2.1918$, $n_{eff-s} = 2.1734$, and $n_{eff-i} = 2.1895$. It is clear that the oscillation period is $1.9\ mm$, and the maxima of the entangled photon pair generation appear at $L = 1\ mm,\ 2.9\ mm,\ 4.8\ mm,\ 6.7\ mm$, and $8.6\ mm$. In addition, we can see that the maximum conversion efficiencies of each peaks increase with the increase of $L$ because of the "quasi-phase matching" between the bound TM mode and the TE modes. The present quasi-phase matching is different from the usual cases. To distinguish this, here we have added quotation marks. There are usually two ways to achieve the quasi-phase matching [76]. One is to control the directions of wave propagation in anisotropic nonlinear crystals. Another is to use periodically poled materials. The present "quasi-phase matching" is obtained by adjusting the waveguide modes and the effective refractive indices at different wavelengths.

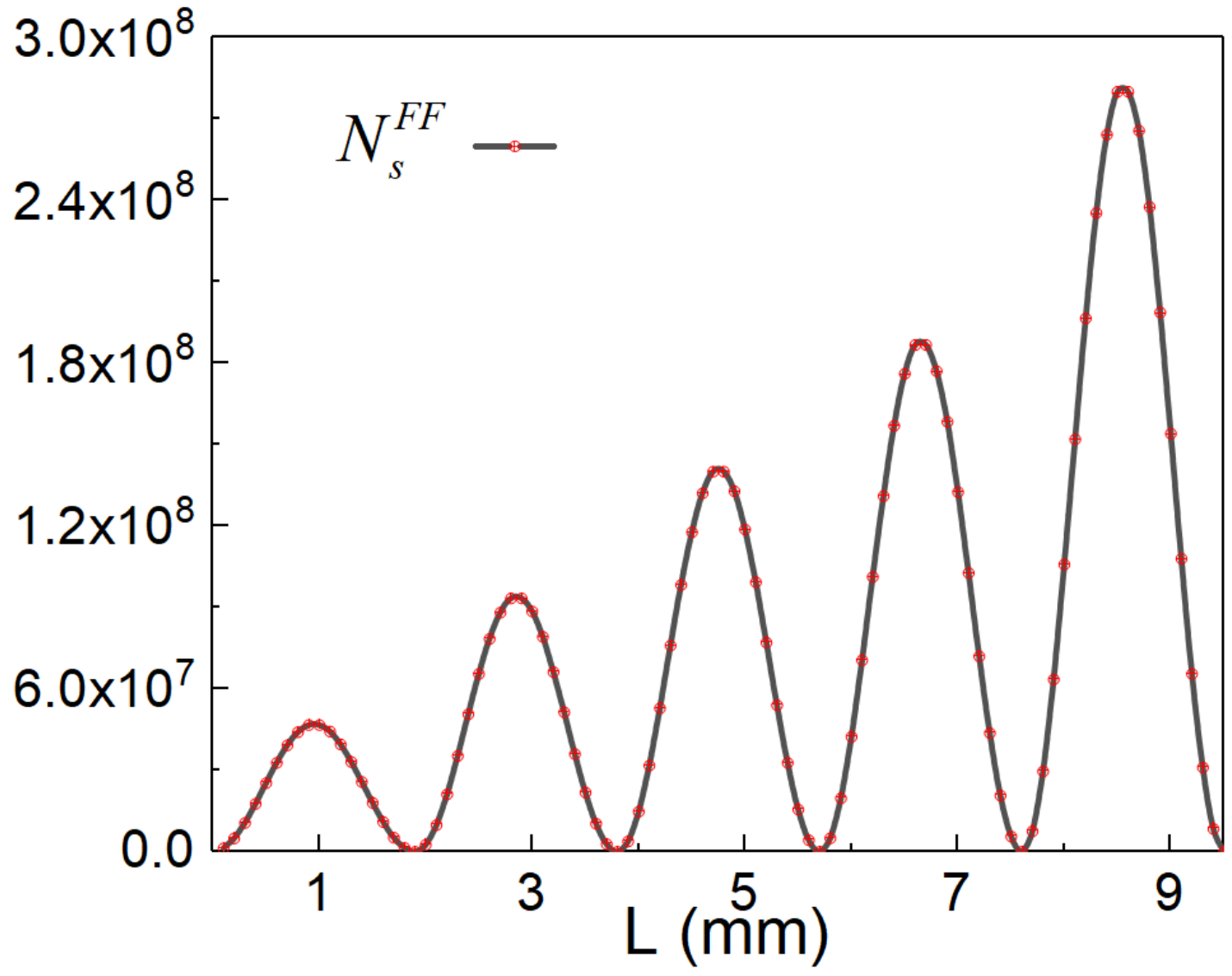

Fig. 2. The variation of the number of coherent photon pairs $N_s^{FF}$ with the waveguide length.

Fig. 3(a) displays the calculated results of the photon counts for the signal and idler fields as functions of the wavelengths $\lambda_s$ and $\lambda_i$ at $L = 1\,mm$. They are also obtained by the self consistent solution of Eq. (23) that makes $\Delta k$ minimum. According to the phase-matching conditions, the signal and idler wavelengths are different. The maximum count of TE-polarized photons appears at $\lambda_s = 1200.5\,nm$, and the corresponding maximum value of TM-polarized photons appears at $\lambda_i = 1400.2\,nm$. This means that polarized entangled photon pairs with different wavelengths have been generated. The optical power of the generated signal depends on the length of the waveguide. The red and black solid lines in Fig. 3(b) display the calculated results at $L = 1\,mm$ and $L = 8.6\,mm$. The generation rate of entangled photon pairs increases linearly with the waveguide length. At $L = 8.6\,mm$, the generation rate is increased by more than five orders of magnitude in contrast to conventional LN waveguides. The high generation efficiency originates from the quasi-phase matching between the TM bound mode and the fundamental TE modes. This can be seen more clearly from the electric field distributions at the corresponding wavelengths shown in Figs. 3(c)-3(h). It is the “quasi-phase matching” between the bound TM mode and the fundamental TE modes that confines the field in the LN waveguide.

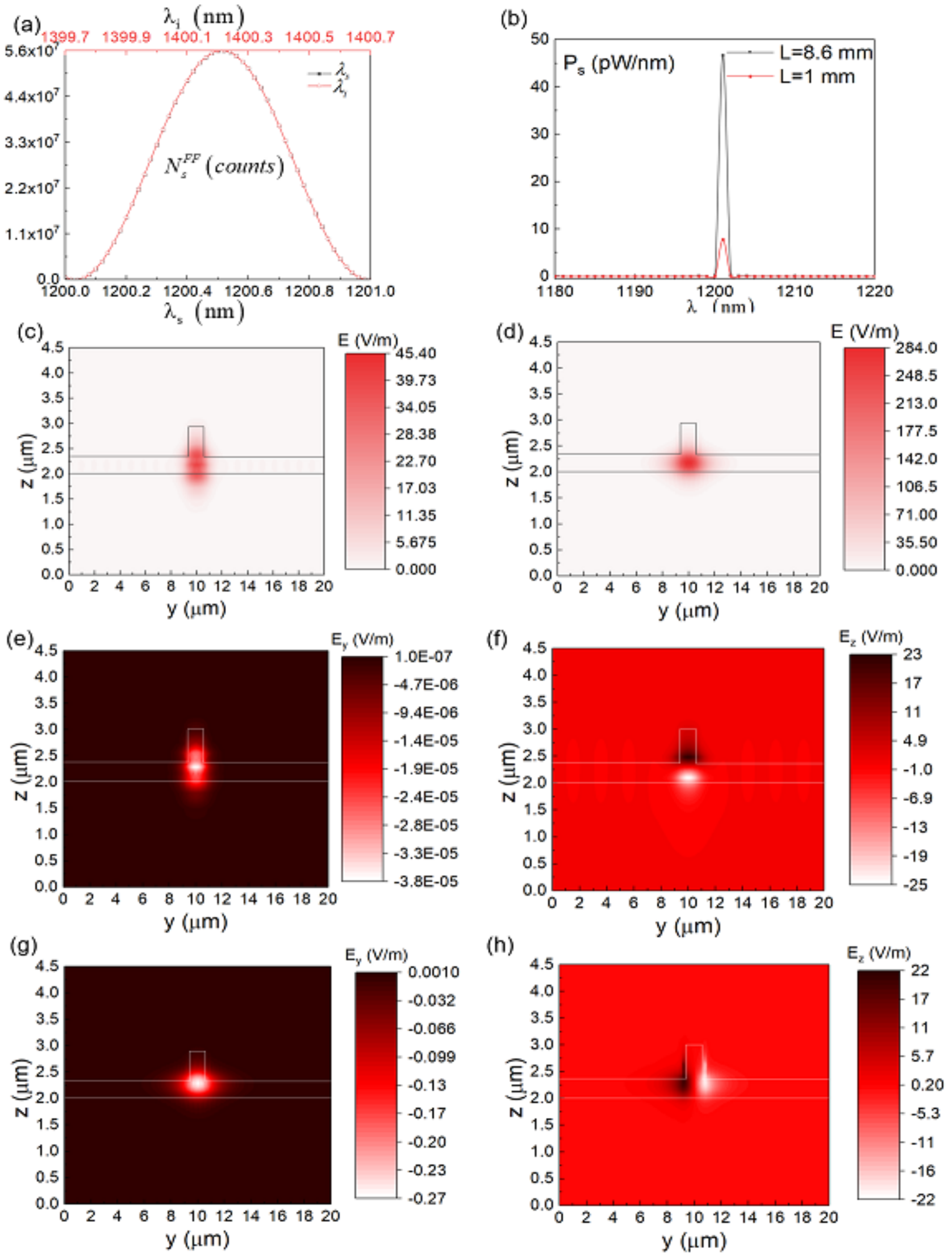

Fig. 3. (a) Photon counts for the signal and idler fields at the pump wavelength $\lambda_p = 633\,nm$, $N_{s,i}^{FF}$ at $\lambda_s = 1200.5\,nm$ and $\lambda_i = 1400.2\,nm$. (b) The signal optical powers from different waveguide lengths under the condition of quasi-phase matching. The red and black solid lines correspond to $L = 1\,mm$ and $L = 8.6\,mm$, respectively. (c)-(h) Electric field distributions at different wavelengths: (c), (e) and (f) Normalization, y component and z component TE electric field at $\lambda_s = 1200\,nm$, (d), (g) and (h) Normalization, y component and z component TM electric field at $\lambda_i = 1399.7\,nm$.

An advantage of such an approach is that polarized entangled photon pairs with different wavelengths can be generated in a wide range of frequencies by choosing the parameters of the structure appropriately. Fig. 4(a) shows the calculated results of the photon counts for the signal and idler fields as functions of the wavelengths $\lambda_s$ and $\lambda_i$ for the pump wavelength $\lambda_p = 1550\ nm$. Here, the thickness of the silica slab is taken as $d = 2.3\ \mu m$, the thickness of the organic waveguide as $h_{wg} = 490\ nm$, the thickness of the LN waveguide as $d_{LN} = 520\ nm$, and the width of the waveguide as $w = 2.7\ \mu m$. The refractive indices are $n_{SiO_2} = 1.44$, $n_{wg} = 1.54$, and $n_{LN-p} = 2.2111$ [69]. The pump power is $P_p = 1mW$. From the quasi-phase matching condition, we obtain $\Delta k = 2.8*10^{-4}\ nm^{-1}$. The first maximum of the entangled photon pair production appears at $L = 3\ mm$. The calculated results in Fig. 4(a) correspond to this case. The maximum count of the TE-polarized photons appears at $\lambda_s = 3000.5\ nm$, and the corresponding maximum value for TM-polarized photons appears at $\lambda_i = 3206.5\ nm$. The optical powers of the generated signals at $L = 3\ mm$ and $L = 10.59\ mm$ are plotted in Fig. 4(b). The generation rate of the entangled photon pairs is improved by more than four orders of magnitude even at $L = 3\ mm$ compared to the conventional case.

The polarized entangled photon pairs can be produced not only in the visible and mid-infrared frequency regions, but can be easily extended to the THz region by designing the waveguide structure appropriately. For example, if the thickness of silica slab is set to $d = 10\ \mu m$, its refractive index should be $n_{SiO_2} = 0.613$ at THz wavelengths. We set the thickness of the organic waveguide as $h_{wg} = 5\ \mu m$ and its refractive index as $n_{wg} = 1.54$ in this case. The thickness of the LN waveguide is $d_{LN} = 3.2\ \mu m$ and its refractive index is taken as $n_{LN-p} = 6.4$ [74], and the width of the waveguide is taken as $w = 69\ \mu m$. The pump power is $P_p = 1mW$. There is material absorption in the LN layer about $20cm^{-1}$. Considering the LN dispersion, its nonlinear coefficient should be $d_{33} = 2200\, pm/V$ [74]. From the quasi-phase matching condition, we obtain $\Delta k = 3.6*10^{-4}\ nm^{-1}$. The first maximum of the entangled photon pair production appears at $L = 4.6\ mm$. At this length, the maximum count of the TE-polarized photons appears at $\lambda_s = 38\ \mu m$, and the corresponding maximum value of the TM-polarized photons appears at $\lambda_i = 42.222\ \mu m$ as shown in Fig. 5(a). The generated optical signal powers at $L = 4.6\ mm$ and $L = 14.28\ mm$ are plotted in Fig. 5(b). The generation rate of the entangled photon pairs is improved by more than four orders of magnitude at $L = 4.6\ mm$ compared with the conventional LN waveguide.

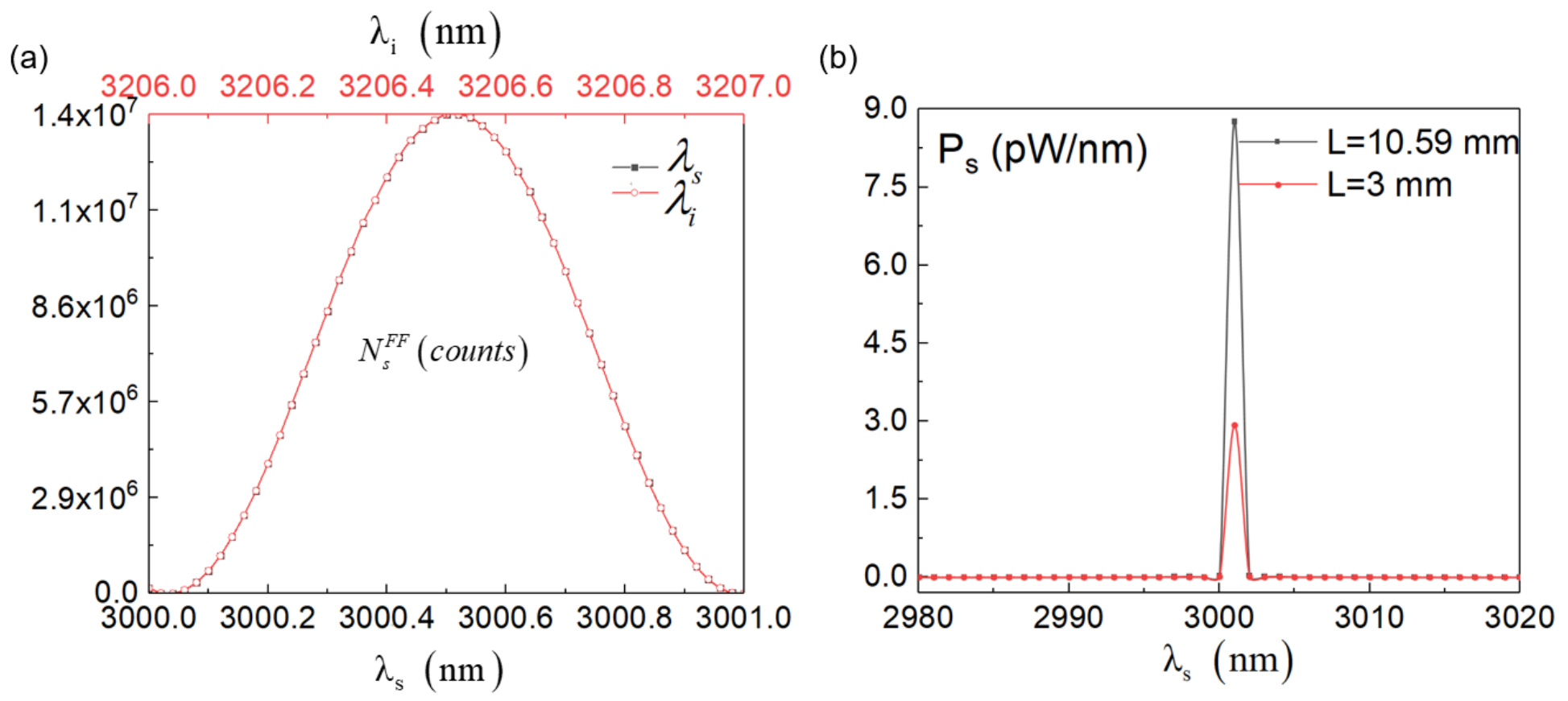

Fig. 4. (a) Photon counts for signal and idler fields at the pump wavelength $\lambda_p = 1550\ nm$, $N_{s,i}^{FF}$ at $\lambda_s = 3000.5\ nm$, and $\lambda_i = 3206.5\ nm$. (b) The signal optical powers from different waveguide lengths under the condition of quasi-phase matching. The red and black solid lines correspond to $L = 3\ mm$ and $L = 10.59\ mm$, respectively.

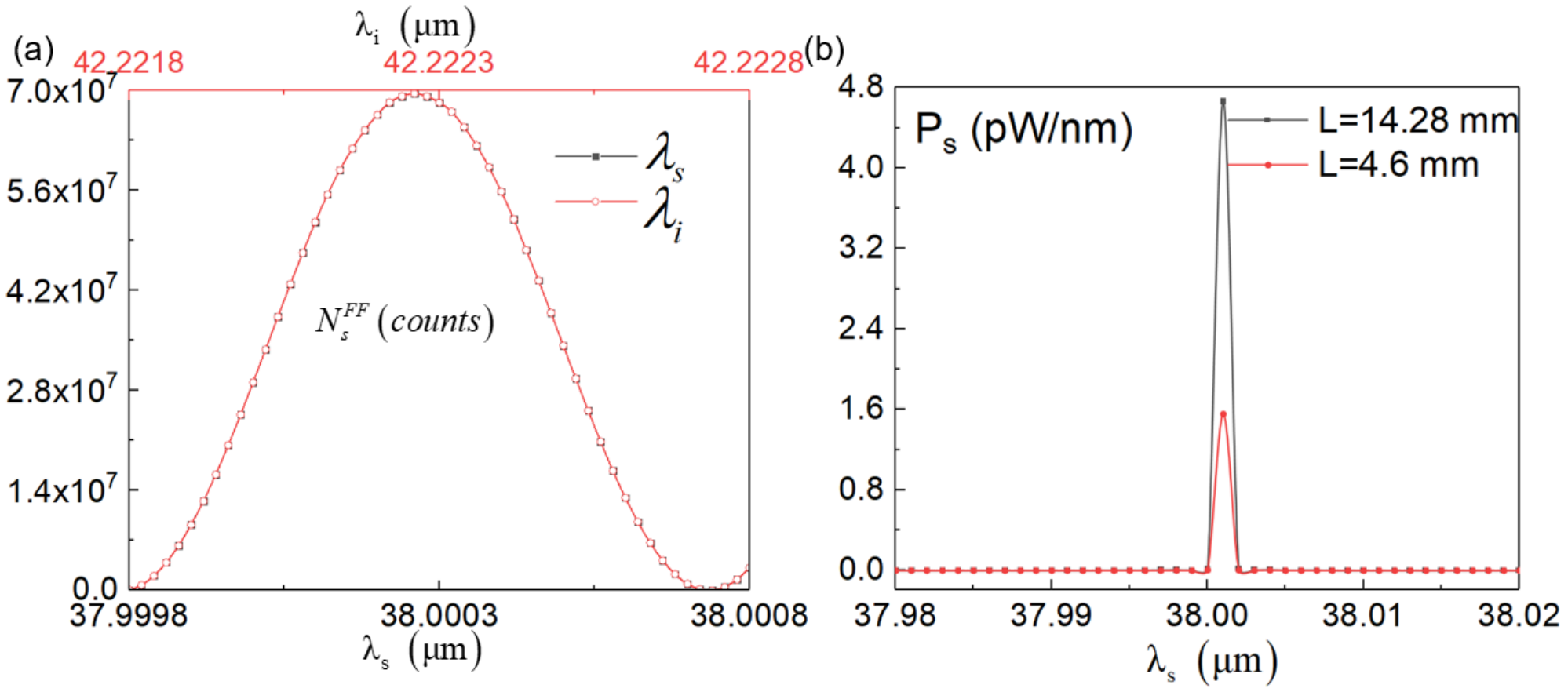

Fig. 5. (a) Photon counts for signal and idler fields at the pump wavelength $\lambda_p = 20\ \mu m$. $N_{s,i}^{FF}$ at $\lambda_s = 38\ \mu m$, and $\lambda_i = 42.222\ \mu m$. (b) The signal optical powers from different waveguide lengths under the condition of quasi-phase matching. The red and black solid lines correspond to $L = 4.6\ mm$ and $L = 14.28\ mm$, respectively.

With the rapid development of fabrication techniques, our theoretical design is easily realized in experiments based on the method described in Refs. [69, 70]. Fig. 6 shows an experimental implementation for generating entangled photon pairs. A high pass filter is used to filter out the pump light, after which the signal light and idler light are separated by a tilted dichromatic mirror. A fiber coupler and a single-mode fiber are used to collect the light for the detectors and a time-to-amplitude converter used for coincidence measurement. Such a scheme is used extensively in experimental measurements for entangled photon pairs [73]. In Ref.[73], it has been demonstrated the generation of entanglement for the signal and idler polarized photons at different wavelengths in periodically poled materials. The phenomena disclosed in the present work are identical with those described in Ref.[73].

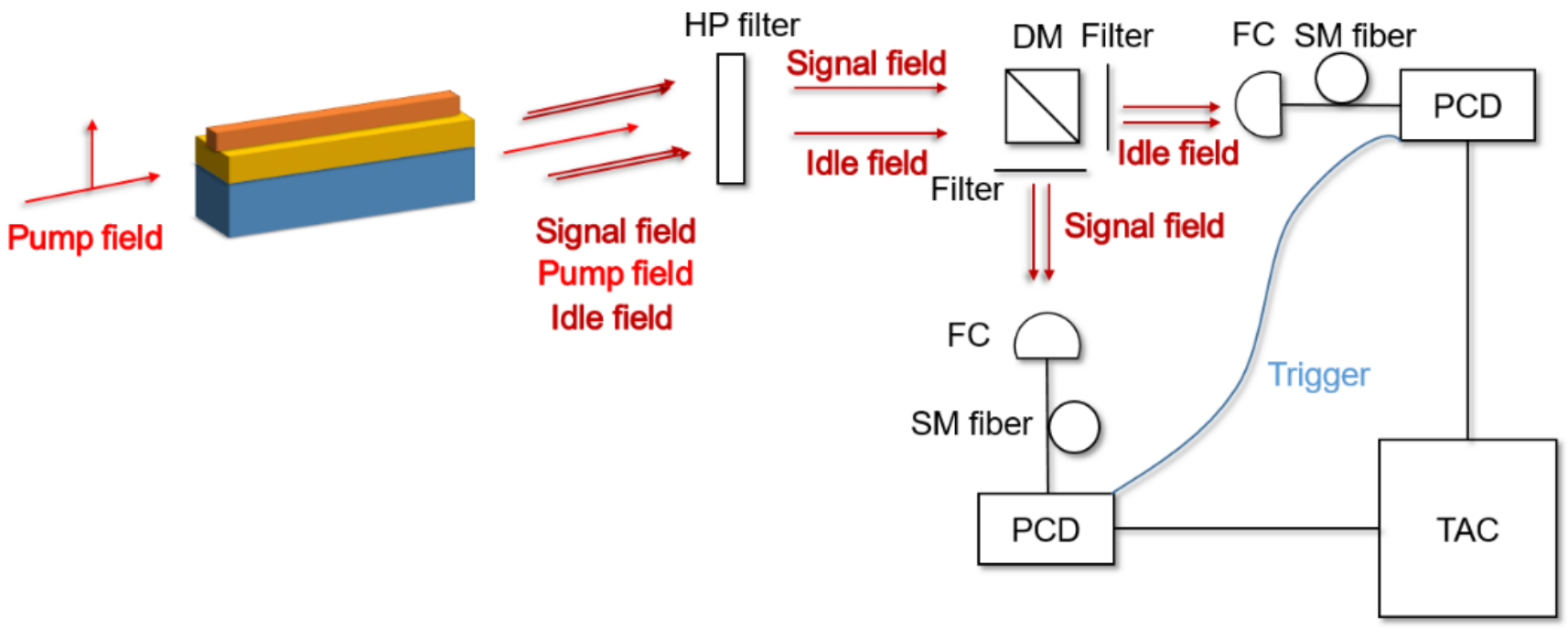

Fig. 6. Experimental measurement scheme for entangled photon generation. HP filter = high pass filter. DM = dichroic mirror. FC = fiber coupler. SM fiber = single mode fiber. PCD = photon-counting detector. TAC = time-to-amplitude converter.

## 4. Conclusion

We proposed a new method to efficiently produce polarization-entangled photon pairs in LN waveguides based on the TM bound state in the TE continuum modes. Differing from periodically-poled LN waveguides, a complicated polarization process is not needed in the design. Compared with the corresponding case without BICs, the generation rate of the entangled photon pair can be improved by several orders of magnitude in properly designed waveguides only a few millimeters long. At the same time, the generation rate increases linearly with the waveguide length. The phenomena can appear in a very wide spectrum range from the visible to THz regions. We believe that the calculated results are of great significance for the development of active integrated quantum chips for various wavelength ranges.

## Funding

National Key R & D Program of China (2017YFA0303800); National Natural Science Foundation of China (91850205).

**Disclosures**

The authors declare no conflicts of interest.

**Appendix**

In the appendix, we give the expressions for $\left(F_{sF}^{(2)}\right)_c$, $\left(F_{iF}^{(2)}\right)_d$ and $\left(F_{pF}^{(2)}\right)_b$, which depend on the boundary conditions and the free field distributions within the layers. The forms of the boundary transition matrices $T_{m,TE}^{(l)}$ and $T_{m,TM}^{(l)}$ are as follows:

$$\begin{cases} T_{m,TE}^{l}\left(\omega_m\right)=\begin{pmatrix} 1+\mu\beta & 1-\mu\beta \\ 1-\mu\beta & 1+\mu\beta \end{pmatrix} \\ T_{m,TM}^{l}\left(\omega_m\right)=\dfrac{1}{2}*\begin{pmatrix} n+\dfrac{\beta}{n} & n-\dfrac{\beta}{n} \\ n-\dfrac{\beta}{n} & n+\dfrac{\beta}{n} \end{pmatrix} \end{cases}, \tag{24}$$

with

$$\begin{cases} \mu=\dfrac{\mu^{(l)}}{\mu^{(l+1)}}\left(\omega_m\right) \\ n=\dfrac{n^{(l)}}{n^{(l+1)}}\left(\omega_m\right) \\ \beta=\dfrac{\beta^{(l)}}{\beta^{(l+1)}}\left(\omega_m\right) \\ \beta=\dfrac{2\pi n}{\lambda} \end{cases}, \tag{25}$$

The free field propagation matrix is:

$$P_m^l\left(\omega_m\right)=\begin{pmatrix} \exp\left(ik_p^l L^l\right) & 0 \\ 0 & \exp\left(-ik_p^l L^l\right) \end{pmatrix}, \tag{26}$$

where $l=1,2,3$ indicates the different materials in the air-LN-air system. Furthermore:

$$F_{mF}^{(2)}(\omega_m) = T_p^{(1)}(\omega_m) * P_m^{(1)}(\omega_m) * T_m^{(0)}(\omega_m) * \begin{pmatrix} \dfrac{1}{(S_m)_{11}(\omega_m)} & -\dfrac{(S_m)_{12}(\omega_m)}{(S_m)_{11}(\omega_m)} \\ 0 & 1 \end{pmatrix}, \quad (27)$$

where

$$S_m(\omega_m) = T_m^3(\omega_m) * \prod_{l=1}^{3} \left[ P_m^l(\omega_m) T_m^{l-1}(\omega_m) \right]. \quad (28)$$